\documentstyle[preprint,aps,eqsecnum,epsfig]{revtex}
\tightenlines
\def\pmb#1{\setbox0=\hbox{#1}%
     \kern-.025em\copy0\kern-\wd0
      \kern.05em\copy0\kern-\wd0
       \kern-.025em\raise.0433em\box0}

\def\beq{\begin{equation}}
\def\eeq{\end{equation}}
\def\bea{\begin{eqnarray}}
\def\eea{\end{eqnarray}}
\begin{document}
\title{\bf Effects of surface vibrations on quadrupole response of 
nuclei}
\author{ V. I. Abrosimov$^{\rm a}$, A.Dellafiore $^{\rm b}$,
F.Matera$^{\rm b}$\\
\it $^{\rm a}$ Institute for Nuclear Research, 03028 Kiev, Ukraine\\
$^{\rm b}$ \it Istituto Nazionale di Fisica Nucleare and Dipartimento
di Fisica,\\
\it Universita' di Firenze, via Sansone 1, 50019 Sesto F.no (Firenze), Italy}
%\date{}
\maketitle
\begin{abstract}
The effect of quadrupole-type surface vibrations on the quadrupole 
response function of heavy nuclei is studied by using a model based on 
the solution of the linearized Vlasov equation with moving-surface 
boundary conditions. By using a separable approximation for the 
residual interaction, an analytical expression is obtained for the 
moving-surface response function. Comparison of the fixed- and 
moving-surface strength functions shows that surface vibrations are 
essential in order to achieve a unified description of the two 
characteristic features of the quadrupole response: the giant resonance and the 
low-lying states. Calculations performed by setting the surface 
tension equal to zero shows that the low-lying strength is 
strongly affected by the surface tension. 

\end{abstract}
\vspace{.5 cm}
\vspace{.5 cm}


\section{INTRODUCTION}

There are two systematic features in the quadrupole excitation 
spectrum of heavy and medium-heavy nuclei: the giant quadrupole resonance
at an energy $\hbar \omega \approx 63 A^{-\frac{1}{3}} {\rm MeV}$ and the 
lower-energy states that have 
often been interpreted as surface vibrations. However the exact nature 
of these states is still under debate since they are 
strongly affected by shell effects and this may be taken as an 
indication that they are not pure surface modes (see e. g. \cite{r&s}, p.14).

There is a vast literature on the quadrupole response of nuclei, here 
we refer only to the nice pedagogical introduction in \cite{b&b} and 
to a recent paper \cite{lac} where the effect of coupling between the 
motion of individual nucleons and surface 
oscillations has been studied in a model that allows also for 
non-linear effects and for collisions between nucleons. The scope of 
our present work is more restricted than that of \cite{lac}, we aim at studying
only the effects of coupling between the 
motion of individual nucleons and surface oscillations of the 
quadrupole type. Our approach is semiclassical and is based on the 
solution of the linearized Vlasov kinetic equation.
Solutions of this equation for finite systems have been obtained by 
using different boundary conditions (fixed- and moving-surface 
\cite{bri,ads}, see also \cite{adm3}), while the 
fixed-surface solution can give a reasonable picture of 
the giant quadrupole resonance, it does not satisfactorily describe also the
low-lying states, however a 
unified description can be achieved within this model
if the moving-surface 
boundary conditions introduced in \cite{ads} are employed. Physically, 
this means that we are including in our model a coupling 
between the motion of individual nucleons and the surface vibrations. 
In our approach this coupling does not
involve  surface vibrations of different multipolarity. Thus, the
present model does not include effects that are not taken 
into account also in the random-phase approximation.

\section{Formalism}
\subsection{ Fixed-surface solution}
Within the fixed-surface theory, and assuming a simplified residual 
interaction of separable form,
\beq
\label{vres}
v(r_{1},r_{2})=\kappa_{2} r_{1}^{2}r_{2}^{2}\,,
\eeq
the quadrupole response function of a spherical nucleus described as 
a system of $A$ interacting nucleons contained within a cavity of radius $R=1.2 
A^{\frac{1}{3}}\: {\rm fm}$ is given by \cite{bur}
\beq
\label{collfix}
{\cal R}_{22}(s)=\frac{{\cal R}_{22}^{0}(s)}{1-\kappa_{2} {\cal R}_{22}^{0}(s)},
\eeq
where $s=\frac{\omega R}{v_{F}}$ is a convenient dimensionless 
variable ($v_{F}$ is the Fermi velocity). The zero-order response 
function ${\cal R}_{22}^{0}(s)$ is analogous to the single-particle 
response function of the quantum theory and is given explicitly by 
\cite{adm2}
\beq
\label{pi0}
{\cal R}_{22}^{0}(s)=
\frac{9A}{8\pi}\frac{1}{\epsilon_{F}}\sum_{n=-\infty}^{+\infty}
\sum_{N=0,\pm2}C_{2N}^{2} \int_{0}^{1} dx x^{2}
~s_{nN}(x){{(Q^{2}_{nN}(x))^{2}}\over {s+i\varepsilon -s_{nN}(x)}}\,.
\eeq
Here $\epsilon_{F}$ is the Fermi energy, the coefficients 
$C_{2N}^{2}$ are $C^{2}_{20}=\frac{1}{4}$ and $C^{2}_{2\pm2}=\frac{3}{8}$.
The functions $s_{nN}(x)$ are defined as
\beq
s_{nN}(x)=\frac{n\pi +N\arcsin(x)}{x}
\eeq
and the quantity $\varepsilon$ is a vanishingly small parameter that 
determines the integration path at poles. The Fourier coefficients
$Q^{2}_{nN}(x)$ are the classical limit of the quantal radial matrix 
elements and are given by:
\beq
Q^{2}_{nN}(x)=(-)^{n} R^{2}\frac{2}{s^2_{nN}(x)}{\Big(}
1 + N \frac{\sqrt{1-x^{2}}}{s_{nN}(x)}{\Big )}\qquad {\rm for}\:(n,N)\neq(0,0)
\eeq
and
\beq
\label{q00}
Q^{2}_{00}(x)=R^{2}(1-\frac{2}{3}x^{2})\,.
\eeq
The response function (\ref{pi0}) involves an infinite sum over $n$, 
however in practice it is sufficient to include only a few terms 
around $n=0$ in order to fulfill the energy-weighted sum rule with 
good accuracy. The form (\ref{q00}) of the coefficient $Q^{2}_{00}$ 
implies that the term $(n,N)=(0,0)$ does not contribute to the 
strength function, hence this term can be omitted from the sum in 
Eq. (\ref{pi0}).

\subsection{ Moving-surface solution}

Within the moving-surface theory of \cite{ads}, the collective response function 
(\ref{collfix}) is replaced by
\beq
\label{rtil}
\tilde{{\cal R}}_{22}(s)=
{\cal R}_{22}(s)+ {\cal S}_{22}(s)\,,
\eeq
with ${\cal R}_{22}(s)$ still 
given by Eq. (\ref{collfix}), while ${\cal S}_{22}(s)$ represents the 
moving-surface contribution. With the simple interaction (\ref{vres})
the function ${\cal S}_{22}(s)$ can be evaluated explicitly as (see 
Appendix)
\beq
\label{surfresp}
{\cal S}_{22}(s)=-\frac{R^{6}}{1-\kappa_{2} {\cal 
R}_{22}^{0}(s)}\: \frac{[\chi^{0}_{2}(s)+\kappa_{2}\varrho_{0}R^{2}
{\cal R}_{22}^{0}(s)]^{2}}{[C_{2}-\chi_{2}(s)][1-\kappa_{2} {\cal 
R}_{22}^{0}(s)]+\kappa_{2}R^{6}[\chi^{0}_{2}(s)+\varrho_{0}R^{2}]^{2}}\,,
\eeq
with $C_{2}= 4\sigma R^{2}$ ($\sigma\approx 1 {\rm MeV\,fm^{-2}}$
is the surface tension parameter obtained from the mass formula) and 
$\varrho_{0}=A/\frac{4\pi}{3}R^{3}$ the equilibrium density.

The functions $\chi^{0}_{2}(s)$ and $\chi_{2}(s)$ are defined as in Refs. 
\cite{adm2} and \cite{adm1} and are given by
\beq
\label{chik}
\chi^{0}_{2}(s) =\frac{9A}{4\pi}\frac{1}{R^{3}}\sum_{n=-\infty}^{+\infty}
\sum_{N=0,\pm2}C_{2N}^{2}\int_{0}^{1} dx x^{2}~
s_{nN}(x){(-)^{n}{Q^{2}_{nN}(x)}
\over {s+i\varepsilon-s_{nN}(x)}},
\eeq
and
\bea
\chi_{2}(s)&=&-\frac{9A}{2\pi}\epsilon_{F}\,(s+i\varepsilon){\Big 
\{}\frac{1}{4}\int_{0}^{1} dx x^{3}\cot[(s+i\varepsilon)x] +\frac{3}{8}
\int_{0}^{1} dx x^{3}{\Big 
(}\cot[(s+i\varepsilon)x-2\arcsin(x)]\nonumber\\
&+& \cot[(s+i\varepsilon)x+2\arcsin(x)]{\Big )}{\Big \}}\,.
\eea

Equation (\ref{surfresp}) is the main result of the present paper, its 
explicit derivation is lengthy but straightforward, the main steps are 
outlined in the Appendix. 
Together with Eq. (\ref{rtil}), Eq. (\ref{surfresp}) gives a unified expression for the 
quadrupole response function, including both the high-energy giant resonance 
and the low-energy excitations. By comparing the two response 
functions (\ref{collfix}) and (\ref{rtil}) we can appreciate the 
effect of the additional surface degree of freedom introduced in 
\cite{ads} and in particular the effect of coupling the motion of nucleons with 
surface vibrations of quadrupole type.

\section{Results}

In Fig.1 we display the strength function ($E=\hbar \omega$)
\beq
S(E)=-\frac{1}{\pi}{\rm Im}\,{\cal R}(E)\,,
\eeq
obtained for $A=208$ using different approximations. The dotted curve 
is obtained from the zero-order response function (\ref{pi0})
and it is similar to the quantum response evaluated in the 
Hartree-Fock approximation. The dashed curve is obtained from the 
collective fixed-surface response function (\ref{collfix}). Comparison 
with the dotted curve clearly shows the effects of collectivity: the 
main strength at about $16\:{\rm MeV}$ is shifted to lower excitation 
energy and the giant quadrupole peak becomes narrower. The strength of 
the interaction (\ref{vres}), chosen in order to reproduce the
experimental value of the giant 
quadrupole resonance energy in $^{208}{\rm Pb}$, is 
$\kappa_{2}=-1.\,10^{-3}\:{\rm MeV\, fm^{-4}}$. this value is close to 
that suggested by the Bohr-Mottelson prescription (\cite{b&m}, p. 
509)
\beq
 \kappa_{2}=-\frac{4\pi}{3}\frac{m \omega_{0}^{2}}{A R^{2}}\approx
 -0.5 \,10^{-3} {\rm MeV\: fm^{-4}}\,,
 \eeq
(with $\omega_{0}= 41 A^{-\frac{1}{3}}\,{\rm MeV}$).

We notice that the width of the giant quadrupole resonance is 
underestimated by the fixed-surface model, this is a well known limit of all 
mean-field calculations that include only Landau damping. Moreover,
there is no sign of a low-energy peak in the fixed-surface response function.
The solid curve instead shows the moving-surface response given by 
Eqs. (\ref{rtil})and (\ref{surfresp}). Now a broad bump appears in the low-energy part of 
the response and a narrower peak is situated at the giant resonance 
energy, thus the moving-surface solution of the Vlasov equation 
introduced in Ref. \cite{ads} accounts for both 
quadrupole modes, although only qualitatively.
Of course the details of the low-energy excitations 
are determined by quantum effects, nonetheless the present semiclassical 
approach does reproduce the average behaviour of this systematic 
feature of the quadrupole response.

Another remarkable feature of the moving-surface response function 
is that now the giant quadrupole peak is narrower than for the 
fixed-surface solution. This is somewhat surprising since we could 
have expected that introducing a further degree of freedom would 
result in a smearing of the peak, however our 
result for the giant resonance 
region is very similar to that of the recent random-pase approximation (RPA) 
calculations of \cite{ham}(cf. Fig.5 of \cite{ham}). Our model does 
include Landau dampig that, however, turns out to be very small in this 
case, clearly some 
additional mechanism is required in order to increase the width of 
the giant resonance. Two such mechanisms have been considered in 
Refs. \cite{bort,lac}, they are the coupling to surface vibrations of 
different multipolarity and the effect of collisions between nucleons. 
It would be interesting to include such effects in the present 
semiclassical theory, however this will be left for future work.

All the strength functions shown in Fig.1 should satisfy the following 
energy-weighted sum rule (EWSR) (see e.g. \cite{b&m}, p. 401):
\beq
\label{ewsr}
\int_{0}^{\infty} dE\,E\,S(E)=\frac{3}{4\pi}\frac{\hbar^{2}}{m}AR^{2}\,.
\eeq
We have numerically checked that, when integrated up to $E=30 {\rm 
MeV}$, the response function shown by the solid curve 
exhausts about $98\%$ of this sum rule. The fraction of EWSR exhausted 
by the dashed and dotted strength functions in the same interval is only $80\%$, 
showing that in these cases there is some more strength at higher energy.

Another interesting moment of the strength function is
the inverse energy-weighted sum rule:
\beq
\label{iewsr}
m_{-1}=\int_{0}^{\infty}dE \frac{S(E)}{E}\,.
\eeq
From the three strength functions shown in Fig.1, we have three 
different inverse moments :
\beq
\label{firstinv}
m_{-1}^{0}=-\frac{1}{2}\lim_{s\to 0}
{\cal R}^{0}_{22}(s)=\frac{139}{140}\frac{1}{16\pi}\frac{AR^{4}}{\epsilon_{F}}
\eeq
(when evaluating $\lim_{s\to 0}{\cal R}^{0}_{22}(s)$
the term $(n,N)=(0,0)$ must be omitted from the sum in Eq. 
(\ref{pi0})),
\beq
\label{secinv}
m_{-1}=m_{-1}^{0} \frac{1}{1+2\kappa_{2}m_{-1}^{0}}
\eeq
and
\beq
\label{thirdinv}
\tilde{m}_{-1}=m_{-1}\Big\{1 + 12 \frac{140}{139} 
\frac{[\frac{17}{20}+2\kappa_{2}m_{-1}^{0}]^{2}}
{[1+2\kappa_{2} m_{-1}^{0}][1+
\frac{8}{3}\frac{4\pi \sigma R^{2}}{A\epsilon_{F}}
+ \frac{27}{50} \frac{140}{139} \kappa_{2} m_{-1}]}\Big\}.
\eeq

The inverse moment of the zero-order strength function (dotted curve 
in Fig.1) $m^{0}_{-1}$ exhausts about $97\%$ of the sum rule 
(\ref{firstinv}) in the range from $0$ to $30\:{\rm MeV}$. The 
collective fixed-surface response (dashed curve in Fig.1) instead 
exhausts almost $99\% $ of the sum rule (\ref{secinv}) in the 
same energy range while the moving-surface strength function exhusts 
almost $100\%$ of the sum rule (\ref{thirdinv}), always in the same 
energy interval. 

It can be seen from Fig.1 that, while the fixed-surface response 
has only one collective pole, the moving-surface quadrupole response 
function displays a two-pole structure. In order to get more 
information about the nature of the 
low-energy peak, we have performed calculations by putting the surface 
tension parameter $\sigma$ equal to zero. The result is shown in 
Fig.2, where the dotted curve corresponds to $\sigma=0$. As expected, 
the giant resonace peak is practically unaffected by the surface 
tension, while the low-energy peak is affected quite substantially. 
The surface tension increases the frequency of the low-enegy peak, 
which, however, is present at a non-vanishing frequency also in the
absence of surface tension. In the opposite limit, if we let 
$\sigma\to\infty $, the fixed-surface response is obtained.

We have performed calculations of the quadrupole response functions 
also for other values of A corresponding to medium-heavy spherical 
nuclei and the results are qualitatively similar to the $A=208$ case, so we do 
not report them here.

A calculation of the isoscalar quadrupole response similar to the 
present one has been made in Ref. \cite{adks}, in that case however a 
rather special external field has been assumed. The external force 
studied in \cite{adks} is a pressure that acts only on the surface of 
the nucleus, perhaps this explains why in that case very little 
strength was found in the region of the giant quadrupole resonance.

\section{Summary and Conclusions}

We have obtained an analytical expression for the isoscalar quadrupole 
response function of nuclei that describes qualitatively the two main 
systematic features of the excitation spectrum: the giant resonance 
and the low-energy states. Our approach is semiclassical and is based 
on the solution of the linearized Vlasov kinetic equation with 
appropriate boundary conditions (moving surface). Comparison of our 
result (full curve) with the quantum response function of Ref. 
\cite{ham} shows that quantum effects modify substantially the 
low-energy region (a discrete state is obtained in \cite{ham} instead 
of our broad bump), while the giant resonance peak is practically the 
same in the quantum and semiclassical approach.

\appendix
\section*{Moving-surface response}

In our semiclassical approach the nucleus is derscribed by means of 
the phase-space density $f({\bf r},{\bf p},t)$. At equilibrium  
the density is $f_{0}({\bf r},{\bf p})=F(h_{0}({\bf r},{\bf 
p}))$, where $h_{0}$ is the equilibrium mean-field hamiltonian. A weak 
external driving field $V_{ext}({\bf r},t)=\beta(t) Q({\bf r})$ 
induces small fluctuations of the equilibrium distribution $f_{0}$ and of 
the mean field. Since the two fluctuations are related, we have a 
typical self-consistency problem. For spherical system the problem can 
be solved by means of the partial-wave expansion \cite{bri,adm3}
\beq
\delta f({\bf r},{\bf p},\omega)=
\sum_{LMN}{\Big[}\delta \tilde{f}^{L+}_{MN}(\epsilon,\lambda,r,\omega)+
\delta \tilde{f}^{L-}_{MN}(\epsilon,\lambda,r,\omega){\Big
]} {\Big (}{\mathcal D}^{L}_{MN}(\alpha,\beta,\gamma){\Big )}^{*}
Y_{LN}(\frac{\pi}{2},\frac{\pi}{2})\,,
\eeq
where ${\mathcal D}^{L}_{MN}(\alpha,\beta,\gamma)$ are the Wigner rotation 
matrices, $\epsilon=h_{0}$ is the particle energy and $\lambda$ its 
angular momentum.

In the approach of \cite{ads} the functions $\delta \tilde{f}^{L\pm}_{MN}$
satisfy the integral equation (see Ref. \cite{adm3} for details)
\bea
\label{flmng}
&&\delta \tilde{f}^{L\pm}_{MN}(\epsilon,\lambda,r,\omega)=
F'(\epsilon)\frac{e^{\pm i[\omega\tau(r)-N\gamma(r)]}}{
\sin[\omega\tau(R)-N\gamma(R)]}\omega p_{r}(R)\delta R_{LM}(\omega)
\\
&&+{\big \{}\int_{r_1}^{r}dr' \tilde{B}^{L\pm}_{MN}(\epsilon,\lambda\, r')
e^{\mp [i\omega\tau(r')-N\gamma(r')]}
+ \tilde{C}^{L}_{MN}(\epsilon,\lambda,\omega){\big \}}
e^{\pm i[\omega\tau(r)-N\gamma (r)]}\nonumber\,,
\eea
with
\beq
\tilde{B}^{L\pm}_{MN}(\epsilon,\lambda,r,\omega)=
F'(\epsilon){\Big (}\frac{\partial}{\partial r}\pm
\frac{iN}{v_{r}(\epsilon,\lambda,r)}\frac{\lambda}{mr^{2}}{\Big)}
{\Big [}\beta(\omega) Q_{LM}(r)+\delta \tilde{V}_{LM}(r,\omega){\Big]}\,,
\eeq
\bea
\label{delvtillm}
&&\delta\tilde V_{LM}(r,\omega)=
\frac{8\pi^{2}}{2L+1}\sum_{N=-L}^{L}
{\Big |}Y_{LN}(\frac{\pi}{2},\frac{\pi}{2}){\Big |}^{2}\\
&&\times \int d\epsilon \int d\lambda \lambda \int \frac{d
r'}{v_{r}(r')}v_{L}(r,r')
{\big [}\delta \tilde f^{L+}_{MN}(\epsilon,\lambda,r',\omega)
+ \delta \tilde f^{L-}_{MN}(\epsilon,\lambda,r',\omega){\big ]}\nonumber
\eea
and
\bea
\label{clmn}
&&\tilde{C}^{L}_{MN}(\epsilon,\lambda,\omega)=
{\Big \{}e^{i2[\omega \tau(R)-N\gamma(R)]}
\int_{r_1}^{R} dr
\tilde{B}^{L+}_{MN}(\epsilon,\lambda,r)
e^{-i[\omega \tau(r)-N\gamma(r)]} \nonumber\\
&&- \int_{r_1}^{R} dr \tilde{B}^{L-}_{MN}(\epsilon,\lambda,r)
e^{i[\omega \tau(r)-N\gamma(r)]}{\Big \}}
{\Big \{}1-e^{i2[\omega
\tau(R)-N\gamma(R)]}{\Big \}}^{-1}\,,\nonumber
\eea
which is equivalent to the linearized Vlasov equation with the 
moving-surface boundary conditions
\beq
\label{msc}
\delta \tilde f^{L+}_{MN}(R)-\delta \tilde f^{L-}_{MN}(R)=2F'(\epsilon)i\omega
p_{r}(R)\delta R_{LM}(\omega)\,.
\eeq

The surface fluctuations $\delta R_{LM}(\omega)$ are related to the functions 
$\delta \tilde{f}^{L\pm}_{MN}$ by
\bea
\label{delrlm}
&&\delta R_{LM}(\omega)=
\frac{8\pi^{2}}{2L+1}\frac{R^{2}}{C_{L}}\sum_{N=-L}^{L}
{\Big |}Y_{LN}(\frac{\pi}{2},\frac{\pi}{2}){\Big |}^{2}\\
&&\int d\epsilon \int d\lambda \lambda p_{r}(R)
{\Big [}\delta \tilde f^{L+}_{MN}(\epsilon,\lambda,R,\omega)+
\delta \tilde f^{L-}_{MN}(\epsilon,\lambda,R,\omega)
-2F'(\epsilon)\delta \tilde V_{LM}(R,\omega){\Big ]}\nonumber\,.
\eea
For a separable interaction of the multipole-multipole type, like 
(\ref{vres}), the 
integral equation (\ref{flmng}) can be reduced to an algebraic equation that, 
in the particular case $L=2$ and $V_{ext}=\beta(t) r^{2}Y_{2M}({\bf 
\hat r})$, gives 
Eqs.(\ref{rtil}) and (\ref{surfresp}) for the quadrupole response 
function.

\begin{figure}
\caption{Quadrupole strength function for a hypothetical nucleus 
of $A=208$ nucleons. The dotted curve shows the zero-order (static 
mean field) aproximation, the dashed curve instead shows the 
collective response evaluated in the fixed-surface approximation. 
The full curve gives the moving-surface response.}
\end{figure}

\begin{figure}
\caption{The full curve is the same as in Fig.1, the dotted curve 
has been obtained in the moving-surface approach for vanishing
surface tension.}
\end{figure}

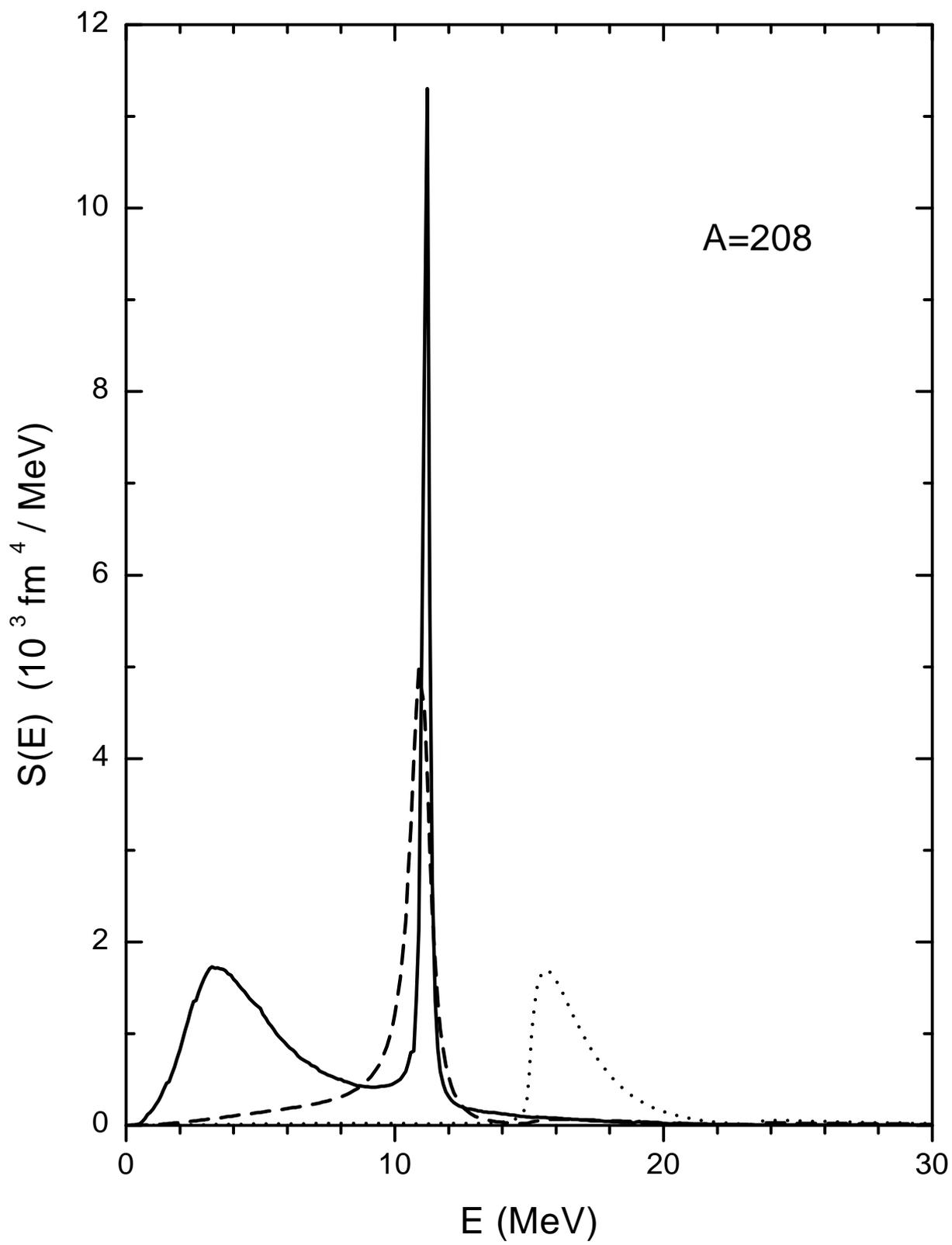
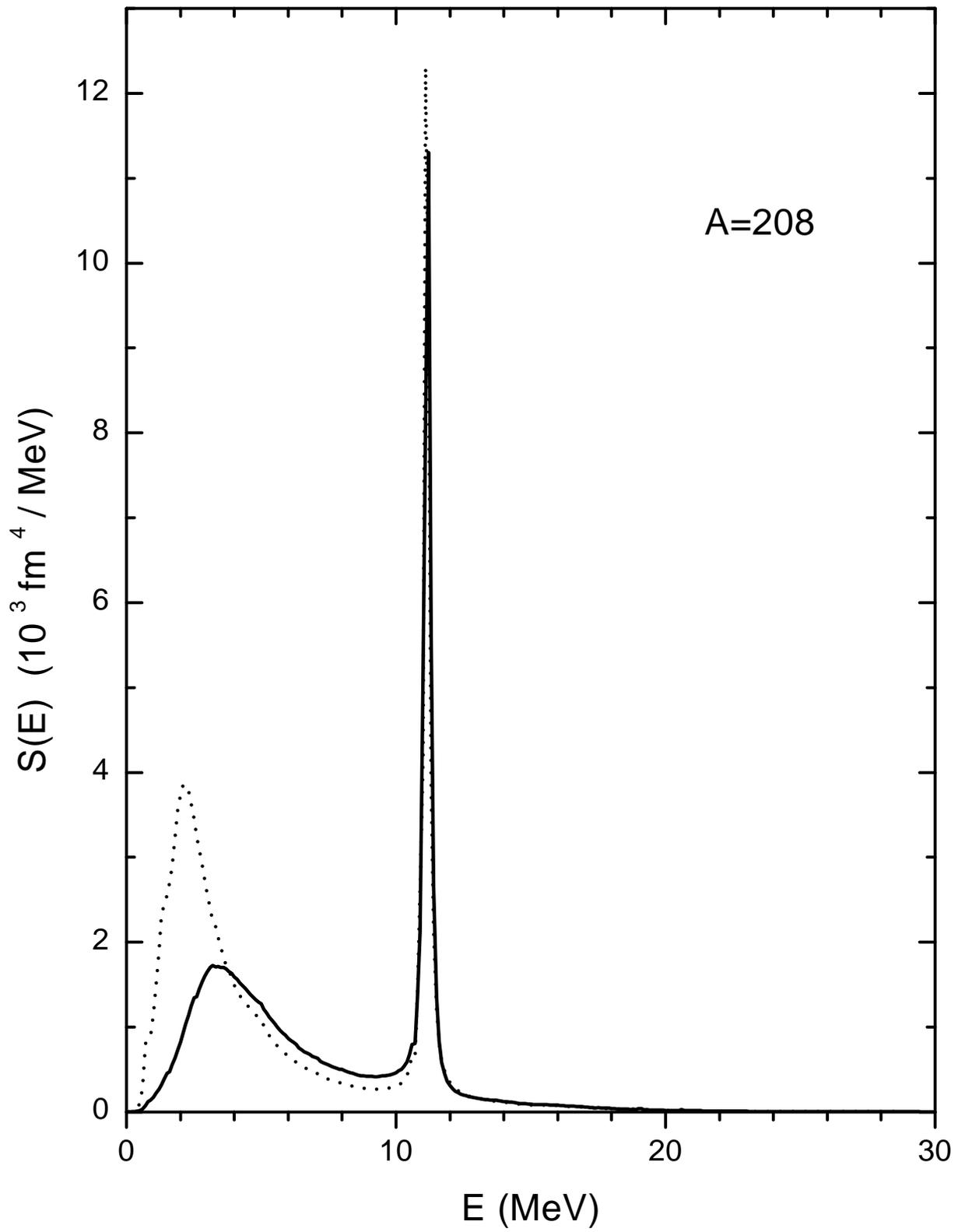
\end{document}